%Paper: hep-ph/9301277
%From: ZHANGX%umdhep.BITNET@VTVM2.CC.VT.EDU
%Date: Tue, 26 Jan 93 13:24 EST

\font\titlefont = cmr10 scaled \magstep2
\magnification=\magstep1
\vsize=20truecm
\voffset=1.75truecm
\hsize=14truecm
\hoffset=1.75truecm
\baselineskip=20pt

\settabs 18 \columns

\def\b{\bigskip}
\def\bb{\bigskip\bigskip}
\def\bbb{\bigskip\bigskip\bigskip}

\def\ce{\centerline}

\def\no{\noindent}

%$$\eqalign{
% put in lines of equations here, each ending in \cr
%}$$

%$$\eqalign{
%put in equations here ending each line with \cr
%} \eqno (1)$$
%the above will put the one between the two lines of equations and set it
%off to the right

% END BEGINNING FORMATS
% BEGIN HEADER

 \rightline{ UMDHEP 93-074}

\bb
\ce{\titlefont {
Operator Analysis for the
Higgs Potential }}
\ce{\titlefont{ and Cosmological Bound on the Higgs-boson Mass}
    \footnote\dag{\rm{ Work supported by a grant from the National
          Science Foundation} }}
\bb
\ce{\bf{ Xinmin Zhang}}

\ce{\it{Center for Theoretical Physics}}
\ce{\it{Department of Physics}}
\ce{\it{University of Maryland}}
\ce{\it{ College Park, MD 20742 }}
\bb
\b
\ce{\bf Abstract}
Using effective field theory at finite temperature, we examine the
impacts of new physics on the electroweak phase transition.
By analysing the high dimensional operators
relevent to the Higgs potential we point out that the Higgs mass bound
obtained by
 avoiding the washout of the baryon asymmetry can be relaxed to the
region allowed by experiments, provided that new physics appears
at the TeV scale.

 \filbreak

It is
generally
realized that baryogenesis
at electroweak scale requires new physics beyond
the standard model since CP violation in the standard model is too small to
yield anything like the observed asymmetry.
Furthermore, in the standard model the upper limit of
the Higgs mass
obtained by avoiding washout of the asymmetry produced during
phase transition lies below the present experimental lower bound.
In dealing with the CP problem,
one possibility to yield sufficient CP
violation is to add high dimensional operators
to the standard model lagrangian\footnote{[F.1]}{Other
possibilities  have also been
 suggested
 such as axion models[1], the singlet majoron models[2],
 the two-Higgs
models[3], the SUSY models[4][5]
 and left-right symmetric models[6].}.
Generally there are
 two operators in the lowest dimension:
$${
O_1 = {\phi^2 \over \Lambda^2} TrW_{\mu\nu} {\tilde W}^{\mu\nu}~~ ,} \eqno(1)$$

$${
O_2 = {\phi^2 \over \Lambda^2} TrG_{\mu\nu} {\tilde G}^{\mu\nu}~~  ,}\eqno(2)$$
\no where $\phi$ is the neutral component
of the Higgs doublet H, which
develops a vacuum expectation value
$( v \simeq 250 GeV )$,
$G_{\mu\nu} ~{\rm and}~W_{\mu\nu}$ are the field strength of
the $SU(3)_c$ and $SU(2)_L$, respectively. The operator
$O_1$ has been examined in Ref.[5] and
$O_2$ in Ref.[7] with simple replacement of the singlet
field by $\phi^2$.
The results show that the observed
 baryon asymmetry can be produced at weak scale
provided that the new physics scale $\Lambda$ is of O(TeV). Moreover,
the predicted values of the electric dipole moments of neutron and electron
 are very close
to the experimental limits[8].
 This approach to electroweak baryogenesis is based on
 the effective
lagrangian method\footnote{[F.2]}{An example of the use of
effective lagrangian to consider the effects of new physics
on electroweak parameters measured at LEP and in low energy experiments
can be found in Ref.[9],
where it is generally assumed that the particle spectrum at low energies
is that of the standard model and new physics preserves baryon and
 lepton symmetry.
},
 namely the effects of new physics being accounted for
by adding to the standard model lagrangian a set of local operators.

In this
brief report we discuss the constraints on the Higgs mass and the
new physics scale due to electroweak baryogenesis.
We will show that the problem of baryon washout can be solved in the
same way as the solution of the problem of sufficient CP violation.
First of all, let us collect some formulas, with which the cosmological
Higgs mass bound is derived in the one-Higgs model.

\no {\bf (a). Baryon number violation rate:}
  The rate of baryon number violation
per unit volume in the
broken phase can be estimated using the formula[10]
$${
{\Gamma}= \kappa T^4 {( {\alpha_W \over {4 \pi} })}^4
              N_{rot}N_{tr} {( { 2 M_W(T) \over {\alpha_W T} })}^7
           exp({- {E_{sph}(T)\over T} }),
}\eqno(3)$$
\no where $\kappa$ is the numerical factor,
$N_{tr}~{\rm and}~ N_{rot}$ the zero mode factors, estimated in Ref.[10].
The factor $E_{sph}(T)$ is the mass of the sphaleron[11]
$${
E_{sph}(T) = {2 M_W(T) \over \alpha_W } A( {\lambda \over g^2} ),
} \eqno(4)$$
\no where $1.5 < A < 2.7$
for $ 0 \le {\lambda \over g^2 } < \infty $.
 For the light Higgs mass we are using, A is very close to
1.5.

\no {\bf (b). Effective potential:}
The temperature dependent gauge boson mass,
 $M_W(T) = {1\over 2}g \phi(T)$, is determined by the effective potential,
which can be expressed
 in the large temperature limit as follows[12]:

$${\eqalign{
V^{eff}_T &=
       D( T^2 - T_0^2 ) \phi^2
       - E T \phi^3 + {\lambda_T \over 4} \phi^4 ,\cr}
}\eqno(5)$$
\no where
$${
D= {1\over 8 v^2} ( 2 M_W^2 + 2 m_t^2 + M_Z^2 )~~;}$$

$${
T_0^2 = {1\over D} ( {m_H^2 \over 4} - 2 B v^2  )
     ~~;}$$

$${
B={3\over {64 \pi^2 v^4}}( 2 M_W^4 + M_Z^4 - 4 m_t^4) ~~; }$$

$${
E= {1\over {6\pi v^3} } ( 2 M_W^3 + M_Z^3 ) ~~;}$$

$${
\lambda_T
= \lambda
- {3 \over {16 \pi^2 v^4} } \biggl(
  2 M_W^4 \ln {M_W^2 \over {\alpha_B T^2} }
      + M_Z^4 \ln {M_Z^2 \over {\alpha_B T^2}}
- 4 m_t^4 \ln {m_t^2 \over {\alpha_F T^2} } \biggr)
{}~~,} $$
\no where
$\ln \alpha_B = 2\ln 4\pi-2\gamma \simeq 3.91$ and
 $ \ln \alpha_F
=2 \ln \pi - 2\gamma \simeq 1.14$.

\no {\bf (c). Higgs mass bound:} In the broken phase,
the vacuum expectation valuve, $\phi(T) \sim
E/ \lambda_T$,  should be big enough to suppresse
the sphaleron rate[13].
For the one-doublet Higgs theory, one has approximately $m_H=
{(2\lambda)}^{1/2}v ~ < ~ 35 GeV$[12].

\b
Now we discuss the corrections of new physics to each
of the above aspects:

\no{\bf (a'). Corrections to baryon number violation rate:}
Since there is still one-doublet
Higgs field in the effective theory, we think that
with Higgs mass in the range of 100 GeV (see below),
``deformed sphaleron" is irrelevent to the calculation of sphaleron rate
\footnote{[F.3]}{ The arguments here are presented in paper[14].
In the one-doublet Higgs model, ``deformed sphaleron"[15] can emerge only
for $m_H > 12 M_W$, which is not physically interesting based on
triviality arguments.}.
So the effect of
high dimensional operators is to modify the coefficient
$A( \lambda / g^2 )$ in eq.(4),
but the effect is generally suppressed by powers of
$\Lambda$.

\no {\bf (b'). Corrections to the effective potential:}
The effect of new physics is to add some high dimensional
operators, contructed out of Higgs field without derivatives, to
the effective potential (5).
The first paper of
Ref.[9] has classified almost
all of the possible operators. From their list, we find one operator
 in the lowest dimension:
$${
 O_3~ =~ \alpha ~ {\phi^6 \over \Lambda^2}~~, }\eqno(6)$$

\no
where $\alpha$ is a free parameter, calculated by matching
the effective theory with the underlying theory.
If we simply add this operator to the effective potential (5),
it violates the renomalization conditions used in Refs.[12][16] to
calculate the effective potential (5),
namely,
 $O_3$
 shifts the vacuum expectation value and Higgs mass
away from that derived from the Higgs
potential (5).
In fact if we fix the vacuum expectation value as it shoud be,
the effects of $O_3$ are renormalizing the Higgs self-couplings.
Following this line of argument,
we redefine a renomalized $O_3$ by
 imposing renormalization conditions[12][16]
 that preserve
the tree level value of $v~ {\rm and}~ m^2_H=2 \lambda v^2$
at
$\phi = v$.
Denoting the renormalized $O_3$ by
$${
V^{(r)}_3 = O_3 + ~counterterms }\eqno(7)$$
\no
where $counterterms
{}~= a ~v^2 ~ \phi^2 ~+~ b ~\phi^4$
with $a, ~b ~\sim O(v^2/ \Lambda^2 )$, using renormalization conditions
[12][16]
\item{i)}
 $ {d \over {d\phi}}  V^{(r)}_3 \bigg|_{\phi =v} = 0 ~ ,$
\b
\item{ii)}
$ {d^2 \over {d\phi}^2}  V^{(r)}_3 \bigg|_{\phi =v} =0 ~ ,$
\b
\no we get
$${
V^{(r)}_3 = \alpha {v^2 \over \Lambda^2} \phi^2 \biggl(- \phi^2
            +v^2 + {1\over 3}{\phi^4 \over v^2} \biggr)~~ .}\eqno(8)$$

There should be two more operators which are not in the list of Ref[9].
They are:

$${
O_4 ~ = ~ \beta ~ {\phi^6 \over \Lambda^2} \ln {\phi^2 \over v^2}
{}~~, }\eqno(9)$$
 $${
O_5 ~ = ~ \delta ~~T\cdot \phi^3 \cdot {\phi^2 \over \Lambda^2}~~, }$$

\no where $\beta$ and
$\delta$ are calculable parameters.
The similarity of these two operators to $O_3$
is that all of them are suppressed by $\Lambda^2$, so they are effects of
new physics. The differences are that,
unlike $O_3$,
$O_4$ is not analytic in $\phi$ and
$O_5$ is very similar to the cubic term in (5). It can be easily proven that
these two operators are generated by
 loops of light particles from high dimensional
operators.

Since $O_4$ and
$O_5$ are doubly suppressed by loop factors and $\Lambda^2$,
practically we can neglect them. So the important correction
of new physics to effective potential is due to $V_3^{(r)}$.

\no {\bf (c'). Corrections to the Higgs mass bound:}
When combined with the discussions above, we conclude that
the most important effect of new physics on the Higgs mass bound is
shifting
$\lambda_T$ in (5) to
$\lambda_T -4 \alpha {v^2\over \Lambda^2}$.
Consequently,
$${
m_H^2 < {( 35 ~GeV)}^2 + 8 \alpha {v^4\over \Lambda^2} ~~.
}\eqno(10)$$
\no Clearly,
we should take $\alpha$ positive. Taking
$\alpha = 1$, we see that the
 Higgs mass can be relaxed to
around 100 GeV if $\Lambda$ is of O(TeV).
 For examples, taking $\Lambda$ to be
3.6TeV, 2.5TeV and 1.9TeV, we get
$M_H < $ 60GeV, 80GeV and 100GeV respectively.

In conclusion,
we have considered the operator-
analysis for electroweak phase transition. Our results show that
electroweak baryogenesis is possible provided that
 new physics is in the TeV range.
However, it should be pointed out that $\Lambda$ cannot be too small,
otherwise $V_3^{(r)}$ will destroy vacuum stability
at zero temperature, since it
 increases the true vacuum energy relative to the false vacuum
energy.
As an example, taking $m_t = 150 GeV$ and
$m_H = 100 GeV$, we find that
$\Lambda$ cannot be lower than about 1 TeV.
Certainly, if $\Lambda$ is too small, our effective lagrangian becomes
unreliable. Instead, we should work on the underlying theory.

We should also mention that the mechanism for
relaxing the Higgs mass bound proposed here is the same as that of Hall and
Anderson[16]; however, the models used
are quite different. In Ref.[16],
 explicit
gauge singlet fields with
invariant bare masses are used,
and in the large mass limit
a operator similar
to $V_3^{(r)}$ is also obtained.
This is understandable, since the effective operator $O_3$ should be
the loop effects of heavy particles,
which include the singlet scalars used in Ref.[16].
 The method used in this paper is more general, and it is
consistent with the approach to electroweak baryogenesis used
in Refs.[5][7].
It should be pointed out that in our discussions we
have neglected the effect of the
running
coupling constants from $\Lambda$ down to T.
This will give rise to some uncertainties in the
determination of $m_H ~{\rm and}~ \Lambda$. However, this uncertainty
 may not be bigger than
that due to the well-known uncertainties in the B-violation rates.

\bbb
I would like to thank R.N. Mohapatra and S. Nussinov
for discussions, and
S. Lee, B-L Young for discussing
the properties associated with sphalerons in effective lagrangian.
\bb

\bb
\ce{\bf References}
\b
\item{[1]} L. McLerran, Phys. Rev. Lett. 62, 1075 (1989);
 V. A. Kuzmin, M. E. Shaposhnikov and I.I. Tkachev,
          Phys. Rev. $\bf 45D$, 466 (1992).

\item{[2]} A.G. Cohen, D.B. Kaplan and A.E. Nelson, Phys. Lett.
           $\bf 245B$, 561 (1990);
          Nucl. Phys. $\bf 349B$, 727 (1991).

\item{[3]} N. Turok and J. Zadrozny, Phys. Rev. Lett.
          $\bf V65$, 2331 (1990); Nucl. Phys. $\bf B358$, 471 (1991);
 L. McLerran, M. Shaposhnikov, N. Turok and M. Voloshin,
          Phys. Lett. $\bf 256B$, 451 (1991).

\item{[4]} A.G. Cohen, D.B. Kaplan and A.E. Nelson, Phys. Lett.
$\bf 263B$, 86 (1991); A.E. Nelson, D.B. Kaplan and A.G. Cohen,
Nucl. Phys. $\bf 373B$, 453 (1992).

\item{[5]} M. Dine, P. Huet, R. Singleton and L. Susskind, Phys. Lett.
           $\bf 257B$, 351 (1991).

\item{[6]} R.N. Mohapatra and X. Zhang, Maryland preprint UMDHEP 92-230
(1992) To appear in Phys. Rev. D.

\item{[7]} R.N. Mohapatra and X. Zhang, Phys. Rev. $\bf 45D$,
           2699 (1992).

\item{[8]} X. Zhang, Maryland preprint UMDHEP 93-07, July 1992.

\item{[9]} For examples, see W. Buchm\"uller and D. Wyler, Nucl. Phys.
          $\bf 268B$, 621 (1986); B. Grinstein and M.B. Wise,
Phys. Lett. $\bf 265B$, 326 (1991); A. de R\'ujula, B. Gavela, O. Pe\'ne
and F.J. Vegas, Nucl. Phys. $\bf 357B$, 311 (1990).

\item{[10]}P. Arnold and L. McLerran, Phys. Rev. $\bf 36D$,
            581 (1987); L. Carson, Xu Li, L. McLerran and R.T. Wang,
             Phys. Rev. $\bf 42D$, 2127 (1990).

\item{[11]} F.R. Klinkhamer and N.S. Manton, Phys. Rev. $\bf 30D$,
2212 (1984).

\item{[12]} M. Dine, R. Leigh, P. Huet, A. Linde and D. Linde,
Stanford university preprint SU-ITP-92-7 (1992) and references therein.

\item{[13]}  M.E. Shaposhnikov, JETP Lett. $\bf 44$,
465 (1986); A.I. Bochkarev, S. Yu. Khlebnikov and M.E. Shaposhnikov,
 Nucl. Phys. $\bf 329B$, 493 (1990), M. Dine, P. Huet and R. Singleton,
            Nucl. Phys. $\bf 375B$, 625 (1992); B.H. Liu,
 L. McLerran and N. Turok, TPI-MINN-92/18-T, April 1992.

\item{[14]} B. Kastening, R.D. Peccei and X. Zhang, Phys. Lett.
     $\bf 266B$, 413 (1991).

\item{[15]}L.G. Yaffe, Phys. Rev. $\bf D40$, 3463 (1989);
J. Kunz and Y. Brihaye, Phys. Lett. $\bf B216$, 353 (1989);
F.R. Klinkhamer, Phys. Lett. $\bf B236$, 187 (1990).

\item{[16]}G. Anderson and L. Hall, Phys. Rev. $\bf D45$, 2685 (1992).

\bye